\documentclass[12pt]{article}
\usepackage{hyperref}
\usepackage{amsmath}
\usepackage{graphicx}
\usepackage{amssymb} 
\usepackage{soul} 
 \usepackage{xcolor}
\definecolor{myblue}{rgb}{0.14,0.11,0.49}
\definecolor{myred}{rgb}{0.74,0.22,0.15}
\definecolor{mygreen}{rgb}{0.05,0.52,0.42}
\definecolor{myyellow}{rgb}{0.96,0.92,0.13}
\definecolor{myorange}{rgb}{1,0.61,0.36}
\definecolor{mypurple}{rgb}{0.71,0.02,1}
\definecolor{noir}{gray}{0.} 

\newcommand{\Mat}[1]{{{\boldsymbol{#1}}}}
\newcommand{\abs}[1]{\left\vert#1\right\vert}
\def\be{\begin{equation}}
\def\ee{\end{equation}}
\def\bea{\begin{eqnarray}}
\def\eea{\end{eqnarray}}
\def\bi{\begin{itemize}}
\def\ei{\end{itemize}}
\def\noi{\noindent}
\def\dd{\mathrm{d}}

\date{}

\title{A simpler solution of the non-uniqueness problem of the covariant Dirac theory}

\author{Mayeul Arminjon\\
\small\it Laboratory ``Soils, Solids, Structures, Risks'', 3SR\\ \small\it (CNRS and Universit\'es de Grenoble: UJF, Grenoble-INP),\\\small\it BP 53, F-38041 Grenoble cedex 9, France.}

\begin{document}
\maketitle

\begin{abstract} 

\noindent Although the standard generally-covariant Dirac equation is unique in a topologically simple spacetime, it has been shown that it leads to non-uniqueness problems for the Hamiltonian and energy operators, including the non-uniqueness of the energy spectrum. These problems should be solved by restricting the choice of the Dirac gamma field in a consistent way. Recently, we proposed to impose the value of the rotation rate of the tetrad field. This is not necessarily easy to implement and works only in a given reference frame. Here, we propose that the gamma field should change only by constant gauge transformations. To get that situation, we are naturally led to assume that the metric can be put in a space-isotropic diagonal form. When this is the case, it distinguishes a preferred reference frame. We show that by defining the gamma field from the ``diagonal tetrad" in a chart in which the metric has that form, the uniqueness problems are solved at once for all reference frames. We discuss the physical relevance of the metric considered and our restriction to first-quantized theory.

\end{abstract}

\section{Introduction and summary}\label{Introduction}

Dirac's original equation, which describes the behaviour of elementary spin half particles, is valid only in Cartesian coordinates on a Minkowski spacetime. It thus has to be modified in order to be written in a curved spacetime, or already in a flat spacetime with non-Cartesian coordinates, e.g. in a rotating frame, as is the case of our Earth. This leads to redefining the Dirac matrices $\gamma ^\mu $ as a {\it field} depending on the position $X$ in the spacetime V. That field is a priori only assigned to satisfy the basic anticommutation relation
\be \label{Clifford}
\gamma ^\mu \gamma ^\nu + \gamma ^\nu \gamma ^\mu = 2g^{\mu \nu}\,{\bf 1}_4, \quad \mu ,\nu \in \{0,...,3\}.
\ee
Here, $(g^{\mu \nu})$ is the inverse matrix of the matrix $(g_{\mu \nu})$ made with the components $g_{\mu \nu}$ of the metric $\Mat{g}$ in a local coordinate system or chart $\chi : X\mapsto (x^\mu )$. 
\footnote{\label{gamma}\ 
The Dirac matrices $\gamma ^\mu $ depend on the chart as is obvious from Eq. (\ref{Clifford}), and in general a chart is defined only locally, i.e. in some open subset U of the spacetime V. However, the $\gamma ^\mu $ matrices are made with the components of the chart-independent ``$\gamma $ field". The latter is a global section of the vector bundle $\mathrm{TV} \otimes  {\sf E}\otimes {\sf E}^\circ$, with ${\sf E}$ the spinor bundle (Sect. \ref{equivalent operators}) and  ${\sf E}^\circ$ its dual vector bundle  \cite{A45}. Using the coordinate basis $(\partial_\mu)$ associated with a chart, a local frame field $(e_a)$ on ${\sf E}$, and its dual frame field $(\theta ^a)$ on ${\sf E}^\circ$, the local expression of the $\gamma $ field is $\gamma = \gamma^{\mu a } _b\ \partial_\mu \otimes e_a \otimes \theta^b$. The Dirac matrices are $\gamma ^\mu \equiv (\gamma^{\mu a } _b)$. Considering the $\gamma $ field leads one immediately to the correct transformation law of the $\gamma ^\mu $ field \cite{A45}.
}
With respect to the original (Dirac's) Dirac equation, we have thus the new possibility that now different admissible fields of Dirac matrices, say $\gamma ^\mu$ and $\widetilde{\gamma} ^\mu$, are related together through a {\it point-dependent} invertible complex matrix, which is unique up to a complex scalar $\lambda (X)$ \cite{Pauli1936,A40}:
\be \label{similarity-gamma}
\widetilde{\gamma} ^\mu =  S^{-1}\gamma ^\mu S, \quad \mu =0,...,3.
\ee
The point-dependent similarity transformation $S$ \cite{BrillWheeler1957+Corr} is usually called a ``local" similarity transformation (e.g. \cite{CabosShishkin1992}). It should not alter the physical properties of the fermions described by the modified Dirac equation, and is thus a {\it local gauge transformation}. The standard modification of the Dirac equation for a curved spacetime is due to Fock and to Weyl; we call it the Dirac-Fock-Weyl (DFW) equation so as to distinguish it from both the original Dirac equation and the alternative modifications of the latter, which were proposed in previous works \cite{A45,A39}. For the DFW equation, the local gauge transformation $S$ is deduced from a local Lorentz transformation applied to a tetrad field. This implies that, for DFW, $S(X)$ at each spacetime point $X \in \mathrm{V}$ belongs to the spin group ${\sf Spin(1,3)}$. The DFW equation is covariant under any such transformation, provided that it be smooth \cite{BrillWheeler1957+Corr, Isham1978, ChapmanLeiter1976}. (This is in addition to being covariant under a general coordinate change, or ``generally-covariant".) Thus, the DFW equation is unique in a given curved spacetime, at least \cite{Isham1978} if the latter is topologically simple.\\

Nevertheless, it has been recently discovered \cite{A43} that, in any given reference frame (see below), or even in any given coordinate system, neither the Hamiltonian operator $\mathrm{H}$ nor the energy operator $\mathrm{E}$ is unique for the DFW equation, and instead both $\mathrm{H}$ and $\mathrm{E}$ depend on the choice of the tetrad field, in such a way that even the energy spectrum is not unique. It had been previously observed on a particular case by Ryder \cite{Ryder2008} that the presence or absence of Mashhoon's ``spin-rotation coupling" term \cite{Mashhoon1988,HehlNi1990} in the DFW Hamiltonian may depend on the choice of the tetrad field. Thus, the gauge invariance of the DFW Lagrangian, that is its invariance under the spin transformations $S: \mathrm{V}\rightarrow  {\sf Spin(1,3)}$, is not a sufficient protection against an unphysical dependence (e.g., of the energy spectrum) on the arbitrary choice between equally valid fields of Dirac matrices. To solve this non-uniqueness problem, we have to restrict the choice of the tetrad field. 
The generic non-uniqueness of the Dirac Hamiltonian and energy operators is true also \cite{A43} for the alternative generally-covariant Dirac equations \cite{A45,A39}, for which the $\gamma ^\mu $ field need not be deduced from a tetrad field. The standard ``Dirac Lagrangian" valid for DFW can be extended to these alternative equations, see Eq. (\ref{Lagrangian}) below, and is invariant under those local gauge transformations $S: \mathrm{V}\rightarrow  {\sf GL(4,C)}$ which satisfy a certain PDE \cite{A43,A42}. However, in the present work, we define the $\gamma ^\mu $ field from a tetrad field, not only for DFW but also for that alternative equation which we consider. Thus, the local gauge transformations are restricted to ${\sf Spin(1,3)}$---see Eq. (\ref{L=Lambda o S}) below. \\

In a recent work \cite{A47}, a solution of this non-uniqueness problem has been proposed, in particular for DFW, based on the following observations. 
\footnote{\
The work \cite{A43} discusses this problem in detail. The work \cite{A47} begins with a summary of the problem and its cause, and analyzes an attempt \cite{GorbatenkoNeznamov2011,GorbatenkoNeznamov2011b} at solving the non-uniqueness problem of the DFW Hamiltonian operator: in a given coordinate chart, there are different ``Schwinger tetrads", related two-by-two by, in general, a time-dependent rotation. \{This can be seen already for a Cartesian chart in a Minkowski spacetime, see Eq. (96) in \cite{A47}---Eq. (94) in the arXiv version.\} That gives by Eq. (15) in \cite{A47} a time-dependent gauge transformation, whence two physically non-equivalent Hamiltonians, Eq. (17) in \cite{A47}. 
Moreover, the non-uniqueness generally subsists even if a unique tetrad field is fixed in a given chart: see Sect. \ref{Pb Cholesky} below, and see point ({\bf iii}) in Sect. \ref{Discussion}.
}
The Hamiltonian and energy operators are defined with respect to a given reference frame \cite{A42}---the latter notion being formally defined as an equivalence class F of charts, any two of which exchange by a purely spatial coordinate change \cite{A44}. The data of a reference 
frame F fixes a unique four-velocity field $v$ \cite{Cattaneo1958,A44}, and also fixes a unique rotation rate field $\Mat{\Omega }$ \cite{Cattaneo1958,Weyssenhoff1937,A47}. It is natural to impose on the tetrad field $(u_\alpha )$ ($\alpha =0, ...,3$) the condition that the time-like vector of the tetrad is the four-velocity of the reference frame: $u_0=v$ \cite{MashhoonMuench2002,MalufFariaUlhoa2007,A47}. It was shown \cite{A47} that then the spatial triad $(u_p)$ ($p=1,2,3$) can only be rotating w.r.t. the reference frame. All tetrad fields $(u_\alpha )$, which have the same time-like vector field $u_0$, and for which the rotation rate tensor field $\Mat{\Xi}$ of the spatial triad $(u_p)$ is the same, give rise to equivalent Hamiltonian operators as well as to equivalent energy operators \cite{A47}. Two natural ways to fix the tensor field $\Mat{\Xi}$ are: i) $\Mat{\Xi}=\Mat{\Omega }$, where $\Mat{\Omega }$ is the unique rotation rate field of the given reference frame, and ii) $\Mat{\Xi}=\Mat{0}$. Either choice, i) or ii), thus provides a solution to the non-uniqueness problem. These two solutions are not equivalent, so that experiments would be required to decide between the two. The difference between the two solutions is likely to be of the order of the Mashhoon spin-rotation coupling term \cite{Ryder2008,A47}, which currently is still too small to be tested. It may also be difficult to implement these solutions in practice, especially the solution i) which involves the tensor {\it field} $\Mat{\Omega }$. Moreover, each solution is valid {\it only in a given reference frame} \cite{A47}.\\

The aim of the present paper is to present a simpler solution of the non-uniqueness problem of the generally-covariant Dirac theory. This solution consists in restricting the choice of the tetrad field to a class, inside which any two choices are related together by a {\it constant} Lorentz transformation of the first tetrad field, which translates into a constant gauge transformation $S$ of the $\gamma ^\mu $ field. This is the closest possible solution to the situation for the original Dirac equation in a Minkowski spacetime, in which any two sets of Dirac matrices exchange by a constant similarity transformation \cite{Pauli1936,A40}. The idea of restricting oneself to $\gamma ^\mu $ fields which exchange by constant similarity transformations is quite obvious from special relativity, but its application to the general Dirac equation in a curved spacetime is not. First, in Section \ref{equivalent operators} we show from previous work \cite{A43} that this idea can work a priori only for two versions of the generally-covariant Dirac equation: DFW and the so-called ``QRD--0" version \cite{A45}. For each of these two versions, two $\gamma ^\mu $ fields that exchange by a constant gauge transformation give rise to equivalent Hamiltonian operators, as well as to equivalent energy operators, and this {\it in any possible reference frame}. Then in Section \ref{Tetrad construc} we present a simple general prescription to select a tetrad field in a given reference frame. We show in Section \ref{Pb Cholesky} that in general this prescription leaves us with the non-uniqueness problem, even if the metric is diagonal---in which case this prescription coincides with the widely used ``diagonal tetrad" prescription (\ref{diagonal-tetrad}), which is thus non-unique in general. To get it unique leads one naturally, as we show, to consider a {\it space-isotropic} diagonal metric:
\be\label{space-isotropic diagonal}
(g_{\mu \nu })=\mathrm{diag}(f,-h,-h,-h), \qquad f>0,\ h>0.
\ee
For that metric, we prove in Section \ref{f -h -h -h} that the ``diagonal tetrad" prescription is unique. 
\footnote{\
The form (\ref{space-isotropic diagonal}) is well known \cite{Ni72}. It has been investigated previously in the context of the Dirac equation \cite{A38,Obukhov2001}, but this was in the case of a stationary metric, $g_{\mu \nu ,0}=0$ (thus here $f_{,0}=h_{,0}=0$). For a stationary metric, independently of its form, Eq. (\ref{Clifford}) makes it natural to choose time-independent Dirac matrices. With this natural choice for a stationary metric, the non-uniqueness problem discussed here does not exist for the DFW equation \cite{A43,A47}.
}
In fact, that prescription applied to that metric leads exactly to the desired property. Namely, if one applies the ``diagonal tetrad" prescription (\ref{diagonal-tetrad}) in each among two charts belonging to the same reference frame, and in each of which the metric has the form (\ref{space-isotropic diagonal}), then the two tetrad fields obtained thus are related together by a constant Lorentz transformation. It follows that, by choosing the diagonal tetrad in a chart, in which the metric has the form (\ref{space-isotropic diagonal}), we solve the non-uniqueness problem in any reference frame. In Section \ref{Discussion}, we review the procedure and its uniqueness features and we show the physical relevance of the metric (\ref{space-isotropic diagonal}). Appendix \ref{FirstQuantized} explains our restriction to first-quantized Dirac theory.

\section{Conditions for equivalent operators}\label{equivalent operators}

For a wave equation on a spacetime manifold, it is the {\it local expression} of the wave equation which can be rewritten in the Schr\"odinger form:
\be \label{Schrodinger-general}
i \frac{\partial \Psi }{\partial t}= \mathrm{H}\Psi, 
\ee
to get the Hamiltonian operator $\mathrm{H}$ that does not involve time derivatives \cite{A47}. Specifically, the Dirac wave function $\psi $ is a section of the spinor bundle ${\sf E}$. The local expression of the Dirac operator involves choosing a local coordinate system on V: $X\mapsto (x^\mu )$, and a local frame field $(e_a)$ on ${\sf E}$: $X\mapsto (e_a(X))$. Thus, in Eq. (\ref{Schrodinger-general}), $t\equiv x^0/c$ is the coordinate time and $\Psi \equiv (\Psi^a)_{a=1,...,4}$ is a function taking its values in ${\sf C}^4$, made with the components $\Psi^a$ of $\psi $ in the frame field $(e_a)$ on ${\sf E}$. We assume that there exists a global orthonormal tetrad field on the tangent bundle TV. Then both the trivial bundle $\mathrm{V} \times {\sf C}^4$ and the complexified tangent bundle $\mathrm{T}_{\sf C}\mathrm{V}$ are spinor bundles \cite{A45}. Thus we have two different choices for ${\sf E}$: {\bf i}) we can take ${\sf E}=\mathrm{V} \times {\sf C}^4$. This is the ``quadruplet representation of the Dirac field'' ({\it QRD}) \cite{A45}, for which the wave function is a complex four-scalar field. This choice is appropriate, in particular, for DFW \cite{A45}. (The DFW wave function transforms as a scalar on a change of the coordinate system \cite{BrillWheeler1957+Corr,ChapmanLeiter1976,deOliveiraTiomno1962}.) Or instead {\bf ii}), we can take ${\sf E}=\mathrm{T}_{\sf C}\mathrm{V}$. This is the ``tensor representation of the Dirac fields" ({\it TRD}), for which the wave function is a complex vector field \cite{A45,A39}.\\

\noindent The operator $\mathrm{H}$ depends in general on the coordinate system or chart, but (for a given $\gamma $ field, see Note \ref{gamma}), it remains the same for two charts which exchange by a purely spatial change \cite{A42,A47}:
\be\label{purely-spatial-change}
x'^0=x^0,\ x'^j=\varphi ^j((x^k)) \qquad (j,k=1,2,3).
\ee
The relation (\ref{purely-spatial-change}) between two charts is an equivalence relation for charts which are all defined on a given open set $\mathrm{U} \subset \mathrm{V}$. We formally define a {\it reference frame} as an equivalence class F of charts for this relation, for a given open set U (the domain of F)  \cite{A44}. This definition corresponds with the notion of a reference frame as being a fictitious fluid: the world lines 
\be\label{x^j = constant}
x^0 \mathrm{\ variable,\qquad }x^j =\mathrm{constant\ for\ }j=1,2,3
\ee
are the trajectories of the particles constituting the reference fluid \cite{Cattaneo1958}. Our definition assumes, moreover, that the time coordinate: $X \mapsto x^0$, is a fixed function. This is made necessary by the fact that the Hamiltonian operator depends on the choice of the time coordinate. We emphasize that the dependence of the operator $\mathrm{H}$ on the reference frame is necessary and natural: it occurs for any wave equation \cite{A47}. That dependence has thus nothing to do with the dependence of $\mathrm{H}$ on the coefficient field $\gamma ^\mu $,
\footnote{\label{A-coeff field}\
For the alternative versions of the generally-covariant Dirac equation \cite{A45}, the r.h.s. is augmented with a term that involves the hermitizing matrix field $A$ \cite{Pauli1936,A40,A42}:
\be\label{Dirac-general}
\gamma ^\mu D_\mu\Psi = -i m \Psi -\frac{1}{2}A^{-1}(D_\mu (A\gamma ^\mu ))\Psi,
\ee 
in order to ensure the current conservation. That term vanishes for the DFW equation \cite{A42}. The matrix $A$ is determined by the data of $\gamma ^\mu $ only up to a positive scalar field $\lambda (X)$ \cite{A42}. Moreover, $A$ is crucial in the definition of the scalar product \cite{A42,A43}, also for DFW with a general set $(\gamma ^{\natural \alpha })$ in Eq. (\ref{Clifford-flat-2}). Thus, $A$ is an additional coefficient field. Upon a local gauge transformation $S$, the field $A$ changes to $\widetilde{A}\equiv S^\dagger A S$ \cite{A40,A45}.
}
which is the non-uniqueness problem that occurs specifically for the generally-covariant Dirac equation, and which is related to the fact that it is a gauge theory. \\

As shown in detail in Ref. \cite{A45}, the connection $D$ used to define the Dirac operator can be any connection on the spinor bundle ${\sf E}$ (for which there are two different choices, see above). The general expression of the Dirac Hamiltonian \cite{A43} involves the matrices $\Gamma _\mu $ of the connection $D$ and/or the corresponding covariant derivatives acting on $\Psi $: $D_\mu \equiv \partial _\mu +\Gamma _\mu $.  On changing the coefficient fields $\gamma ^\mu $ [and $A$, Note \ref{A-coeff field}] by a local gauge transformation $S$ according to Eq. (\ref{similarity-gamma}), there are two distinct possibilities for the $\Gamma _\mu $ matrices \cite{A46}:
\bi
\item Either [let us call this ``Choice (i)"] they change according to the rule \cite{ChapmanLeiter1976}
\be\label{Gamma-tilde-psitilde=S^-1 psi}
\widetilde{\Gamma }_\mu = S^{-1}\Gamma _\mu S+ S^{-1}(\partial _\mu S),
\ee
which is the necessary and sufficient condition in order that the Dirac equation be covariant under the gauge transformation (\ref{similarity-gamma}), accompanied by the change
\be \label{similarity-psi}
\widetilde{\Psi } =  S^{-1}\Psi
\ee
of the wave function. This rule applies automatically if the transformation (\ref{similarity-gamma}), (\ref{similarity-psi}) is ``passive", i.e., if it results from a mere change of the frame field on the spinor bundle ${\sf E}$, the connection being left unchanged \cite{A45}. On the other hand, consider an ``active" gauge transformation, i.e., one for which Eqs. (\ref{similarity-gamma}) and (\ref{similarity-psi}) relate the expressions of, respectively, two different Dirac $\gamma^\mu $ fields and two different Dirac $\psi $ fields, the frame fields on ${\sf E}$ and (for $\gamma^\mu $) on its dual ${\sf E}^\circ$ and on TV being left unchanged \cite{A45}. In that case, the rule (\ref{Gamma-tilde-psitilde=S^-1 psi}) means that the connection $D$ on ${\sf E}$ is changed for another connection $\widetilde{D}$ on ${\sf E}$. The rule (\ref{Gamma-tilde-psitilde=S^-1 psi}) is verified for DFW by the very construction of DFW \cite{ChapmanLeiter1976}, whether the gauge transformation (\ref{similarity-gamma}), (\ref{similarity-psi}) is passive or active.

\item Or still [``Choice (ii)"], the $\Gamma _\mu $ matrices may be left invariant \cite{A42,A43}, meaning that the connection on the spinor bundle ${\sf E}$ is left unchanged, the frame field being also left unchanged:
\be\label{Gamma-tilde=Gamma}
\widetilde{\Gamma }_\mu = \Gamma _\mu.
\ee
Although this condition does not generally ensure the covariance of the Dirac equation under the gauge transformation (\ref{similarity-gamma}), (\ref{similarity-psi}) \cite{A42}, it does make sense: any two sets of Dirac matrices fields, say $\gamma ^\mu $ and $\widetilde{\gamma }^\mu $, both satisfying the anticommutation relation (\ref{Clifford}), are equally valid coefficient fields for the Dirac equation and thus for the Hamiltonian derived therefrom, whether or not they lead to equivalent Dirac equations. Choice (ii) is appropriate to investigate the alternative versions of the curved-spacetime Dirac equations, because each of them is based on using a specific connection on the relevant bundle ${\sf E}$ \cite{A45,A39}. Thus, each of them has the property that, in contrast with DFW, it does not lead to a unique Dirac equation in a given spacetime $(\mathrm{V},\Mat{g})$ \cite{A42,A45}, so that Choice (i) does not make much sense for these alternative versions.
\ei

In Ref. \cite{A43}, the conditions on the gauge transformation $S$ have been derived in order that the Hamiltonian $\widetilde{\mathrm{H}}$ got after the application of $S$ be physically equivalent to the starting one, i.e., in order that
\be \label{similarity-invariance-H}
\widetilde{\mathrm{H} }  =  S^{-1}\,\mathrm{H}\, S.
\ee
This has been done separately for, on one hand, DFW, giving the very simple condition 
\be\label{partial_0 S=0}
\partial_0 S=0       \qquad (\mathrm{DFW}),
\ee
and for, on the other hand, two alternative versions of the Dirac equation \cite{A39}, based on TRD. It is easy to check that the derivation made for the TRD case, and thus the condition derived for that case \cite{A43}, applies in fact also, as it is, to ``Choice (ii)" above, independently of whether QRD or TRD is being considered. Thus, for Choice (ii), we have the following condition in order that Eq. (\ref{similarity-invariance-H}) be satisfied:
\be\label{H=S^{-1}HS-TRD}
B^0(\partial _0 S)S^{-1} =[B^\mu (D_\mu S)S^{-1}]^a         \qquad (\mathrm{Choice\ (ii)}),
\ee
where $B^\mu \equiv A\gamma ^\mu $, $A$ being the hermitizing matrix field \cite{Pauli1936,A40,A42}, and where $M^a\equiv \frac{1}{2}(M-M^\dagger)$ is the antihermitian part of a matrix $M$, with $M^\dagger\equiv (M^*)^T$ denoting its Hermitian conjugate. Moreover, $D_\mu S$ is the $4\times 4$ matrix made with the covariant derivatives $D_\mu S^a_{\ \, b}$ of the tensor field 
\be\label{S=S^a_b e_a theta^b}
\mathcal{S}=S^a_{\ \, b}\, e_a \otimes \theta ^b,
\ee
where $(e_a)$ is the chosen frame field on ${\sf E}$ and $(\theta ^b)$ is the dual frame field on the dual vector bundle ${\sf E}^\circ$, and is given by \cite{A45}
\be\label{D_mu S}
D_\mu S \equiv \left( D_\mu S^a_{\ \, b} \right)=\partial_\mu S + \Gamma_\mu\,S - S\,\Gamma_\mu.
\ee 
Note that, from (\ref{S=S^a_b e_a theta^b}), it follows that the complex matrix defining a local similarity transformation, $S\equiv (S^a_{\ \, b})$, behaves as a scalar under a coordinate change, the frame field $(e_a)$ on ${\sf E}$ being left unchanged. \\

In the case of time-dependent fields, the Dirac Hamiltonian $\mathrm{H}$ is in general not Hermitian \cite{A42,Leclerc2006}. Then, the relevant spectrum is that of the Hermitian part of $\mathrm{H}$:
\be\label{E:=H^s}
\mathrm{E}=\mathrm{H}^{s}\equiv \frac{1}{2}(\mathrm{H}+\mathrm{H}^\ddagger ),
\ee
where the Hermitian conjugate operator $\mathrm{H}^\ddagger$ is with respect to the unique scalar product identified in Ref. \cite{A42}. The name ``energy operator" for the operator $\mathrm{E}$ is justified, see Ref. \cite{A43} and Eq. (\ref{Field energy = mean value of E}) below. As with Eqs. (\ref{partial_0 S=0}) and (\ref{H=S^{-1}HS-TRD}) for $\mathrm{H}$, in Ref. \cite{A43}, the conditions on the gauge transformation $S$ have been derived in order that the energy operator $\widetilde{\mathrm{E}}$ got after the application of $S$ be physically equivalent to the starting one, i.e., in order that $\widetilde{\mathrm{E} } = S^{-1}\,\mathrm{E}\, S$. These are:
\be\label{SEtilde=ES DFW}
 \left[B^0(\partial _0 S)S^{-1}\right]^a = 0 \qquad (\mathrm{DFW}),
\ee
and
\be\label{SEtilde=ES modified}
\left[B^\mu (D_\mu S)S^{-1}-B^0(\partial _0 S)S^{-1}\right]^a =0 \qquad (\mathrm{Choice\ (ii)}).
\ee

\vspace{5mm}
Hence, for DFW, both the Hamiltonian and the energy operator stay invariant in the sense of Eq. (\ref{similarity-invariance-H}) after a gauge transformation (\ref{similarity-gamma}) applied to the field $\gamma ^\mu $, provided that the gauge transformation $S$ is time-independent, Eqs. (\ref{partial_0 S=0}) and (\ref{SEtilde=ES DFW}). However, the validity of the condition $\partial_0 S=0$ depends on the coordinate system. The chain rule shows that, in order to ensure that it applies in any chart, the stronger condition 
\be\label{partial_mu S=0}
\partial_\mu  S=0       \quad (\mu =0,...,3)
\ee
has to be valid. Thus, for DFW, the wish to get equivalent Hamiltonian operators and equivalent energy operators before and after a gauge transformation, independently of the chart, leads us naturally to impose the restriction that the gauge transformation should be a ``global" one, i.e., $S$ constant over the spacetime. On the other hand, for Choice (ii), the presence of the covariant derivatives $D_\mu S$ in the corresponding Eqs. (\ref{H=S^{-1}HS-TRD}) and (\ref{SEtilde=ES modified}) makes the situation more complex. However, there is one alternative version of the generally-covariant Dirac equation for which it is simple. This is the ``QRD--0" version \cite{A45}, which is got by taking the {\it trivial connection} on the bundle ${\sf E}=\mathrm{V} \times {\sf C}^4$:
\be\label{Trivial connection}
\Gamma _\mu =0 \qquad \mathrm{in\ the\ canonical\ frame\ field\ } (E_a),
\ee  
where the canonical global frame field $X \mapsto (e_a(X))\equiv (E_a)$ on $\mathrm{V} \times{\sf C}^4$ is the constant canonical basis $(E_a)$ of ${\sf C}^4$, i.e., $E_a\equiv (\delta ^b_a)$. Thus, for QRD--0 with the canonical frame field, we have $D_\mu S=\partial _\mu S$ from (\ref{D_mu S}) and (\ref{Trivial connection}). Hence, in that case, it is obvious from Eqs. (\ref{H=S^{-1}HS-TRD}) and (\ref{SEtilde=ES modified}) that $\partial_\mu  S=0$ is a sufficient condition in order that the gauge transformation $S$ lead to a Hamiltonian operator equivalent to the starting one, and the same is true for the energy operator. \\

Therefore, we will try to define a natural class of coefficient fields $\gamma ^\mu $ [and $A$, Note \ref{A-coeff field}], any two of which exchange by a {\it constant} gauge transformation $S$. If we succeed in that, then, in any possible reference frame, one and the same operator $\mathrm{H}$ [in the sense of Eq. (\ref{similarity-invariance-H})] will be defined from any two choices of coefficient fields $\gamma ^\mu $ and $\widetilde{\gamma }^\mu $ taken from that class, as well as one and the same operator $\mathrm{E}$. Moreover, this will be true for both DFW and QRD--0.

\section{Tetrads come from a square root of the metric}\label{Tetrad construc}

In a given chart $\chi :X\mapsto (x^\mu )$, a tetrad field $(u_\alpha)$ is given by its matrix $a\equiv (a^\mu  _{\ \,\alpha})$, such that $u_\alpha = a^\mu _{\ \, \alpha } \partial _\mu $. The orthonormality condition for $(u_\alpha)$ writes then
\be \label{ortho-tetrad}
\Mat{g}(u_\alpha  , u_\beta ) = g_{\mu \nu }\,a^\mu  _{\ \,\alpha}  \,a^\nu _{\ \,\beta}   = \eta _{\alpha \beta},\ i.e.\quad a^T\,G\,a=\eta,
\ee
where $\eta \equiv \mathrm{diag}(1,-1,-1,-1)$ is the Minkowski metric and $G\equiv (g_{\mu \nu })$ is the metric's matrix in the chart $\chi $. Most of the explicit constructions of a tetrad field in a given spacetime concern the case of a ``diagonal metric", i.e., one whose matrix is diagonal in the chart considered:
\be\label{diag G}
G = \mathrm{diag}(d_\mu ),
\ee
that chart being admissible \cite{Cattaneo1958}, i.e., $g_{00}>0$ and the $3 \times 3$ matrix $(g_{jk})\ (j,k=1,2,3)$ is negative definite, thus here $d_0 > 0$  and $d_j<0 \ (j=1,2,3)$. The construction then consists (e.g. Ref. \cite{VillalbaGreiner2001}) in defining the tetrad from the following matrix:
\be\label{diagonal a}
a= (a^\mu _{\ \, \alpha }) \equiv \mathrm{diag}(1/\sqrt{ \abs{d_\mu} } ).
\ee
This defines the ``diagonal tetrad"
\be\label{diagonal-tetrad}
u_\alpha  \equiv \delta _\alpha ^\mu \,\partial _\mu /\sqrt{\abs{d_\mu }}.
\ee
That particular tetrad is orthonormal for the particular metric (\ref{diag G}), because it satisfies the general condition (\ref{ortho-tetrad}).
\\

Now consider a chart in which the metric's matrix $G$ has a general form, though satisfying the admissibility condition for the chart, just recalled. In that general case: i) as for the standard Cholesky factorization, one may show with the Minkowski metric $\eta$ (replacing the identity matrix of the standard Cholesky method) that there exists one and only one solution of the equation
\be\label{Lorentz-Cholesky}
b^T\eta b = G \qquad (G=G(X), X\in \mathrm{V}),
\ee
which is a lower triangular matrix $b=C(X)$ with (strictly) positive diagonal entries. ii) Any other solution $b$ of (\ref{Lorentz-Cholesky}) has necessarily the form
\be
b=L C
\ee
with $L=L(X)$ some Lorentz transformation, $L(X) \in {\sf O(1,3)}$.
\footnote{\label{Reifler}\,
These two precise points were stated to the author by Frank Reifler \cite{Reifler2008}. Point (ii) is easy to check: if $b$ and $b'$ verify (\ref{Lorentz-Cholesky}), then $L\equiv b'b^{-1}$ verifies $L^T\,\eta \,L=\eta $. Point (i) is proved in Ref. \cite{A48v2}.
} \\

Note that any solution $b$ of (\ref{Lorentz-Cholesky}) may be termed a ``square root of the metric", \footnote{\label{OPtheory}
Works by Ogievetski\u i \& Polubarinov \cite{OgievetskiiPolubarinov1965} and Pitts \cite{Pitts2011} are relevant in this connection \cite{A48v2}.
} 
since $b\cdot c \equiv b^T\eta c$ is bilinear. These solutions are in one-to-one correspondence with orthonormal tetrads $(u_\alpha )$ in a metric whose matrix is the general $G$, because Eq. (\ref{Lorentz-Cholesky}) for $b$ is equivalent to the condition (\ref{ortho-tetrad}) for
\be\label{a=binv}
a\equiv b^{-1}.
\ee
The Dirac $\gamma ^\mu $ field associated with a tetrad field is (see e.g. Refs. \cite{BrillWheeler1957+Corr,ChapmanLeiter1976}): 
\be \label{flat-deformed}
 \gamma ^\mu(X) = a^\mu_{\ \,\alpha}(X)  \ \gamma ^{ \natural \alpha},
\ee 
where the constant matrices $\gamma ^{\natural \alpha }$ ($\alpha =0,...,3$) obey the anticommutation relation with the Minkowski metric:
\be \label{Clifford-flat-2}
\gamma ^{\natural \alpha } \gamma ^{\natural \beta  } + \gamma ^{\natural \beta  } \gamma ^{\natural \alpha }   = 2\eta^{\alpha\beta}\,{\bf 1}_4, \quad \alpha ,\beta \in \{0,...,3\}.
\ee
Thus, for a general form of the metric, we are able to define a field of Dirac matrices, by solving Eq. (\ref{Lorentz-Cholesky}) and then by defining the matrix $a$ of an orthonormal tetrad field according to Eq. (\ref{a=binv}). Furthermore, we seem to have a preferred choice for $b$ in Eq. (\ref{a=binv}): namely, $b= C(X)$, the unique Cholesky decomposition of the metric's matrix $G(X)$. Since a diagonal matrix is a fortiori triangular, $C\equiv \mathrm{diag}(\sqrt{ \abs{d_\mu} } )$ is the Cholesky decomposition of the metric (\ref{diag G}). Hence the ``diagonal tetrad" (\ref{diagonal-tetrad}) associated with a diagonal metric, which in our opinion is a natural choice for that case, is a particular case of this choice $b= C$. But, in the general case of a metric that is not diagonal in the chart considered, there does not seem to be any physical reason (as opposed to a merely computational one) to prefer the tetrad field associated with $C(X)$ over any other possible tetrad field. Even apart from this, there is a serious difficulty, which we analyze in the next section.

\section{Non-uniqueness of the ``Cholesky prescription"}\label{Pb Cholesky}

What is physically given is the reference frame (a three-dimensional congruence of time-like world lines), not the coordinate system, for which there is a vast functional space of different choices within a given reference frame. As we recalled in Sect. \ref{equivalent operators}, the Hamiltonian and energy operators $\mathrm{H}$ and $\mathrm{E}$, hence the energy spectrum, depend naturally (for any wave equation) on the reference frame as we define it. That is, $\mathrm{H}$ and $\mathrm{E}$ depend naturally on the equivalence class modulo the purely spatial coordinate changes (\ref{purely-spatial-change}). Unfortunately, for the curved-spacetime Dirac equations, the operator $\mathrm{H}$ (as well as the operator $\mathrm{E}$) is not unique even in a given chart, hence it is a fortiori non-unique in a given reference frame. What we are trying to do, is to find a particular prescription---a particular way of choosing the gamma field in a given chart---ensuring the physical uniqueness of each of these operators. Thus, if we have two admissible charts $\chi $ and $\chi '$ that belong to the same reference frame, applying a such prescription successively in the chart $\chi $ and in the chart $\chi '$ should provide operators $\mathrm{H}$ and $\mathrm{E}$, then $\mathrm{H}$$'$ and $\mathrm{E}$$'$, with $\mathrm{H}$$'$ (respectively $\mathrm{E}$$'$) being physically equivalent to $\mathrm{H}$ (respectively $\mathrm{E}$). Setting
\be
P^\mu _{\ \,\nu  }\equiv \frac{\partial x^\mu }{\partial x'^\nu },
\ee
we have from (\ref{purely-spatial-change}):
\be\label{P = 1 bloc 3 x 3}
P^0 _{\ \,0  }=1, \qquad P^0 _{\ \,j  }=P^j _{\ \,0  }=0 \ (j=1,2,3).
\ee
Let us calculate the local transformation $L$ that transforms a first tetrad $(u_\alpha )$, having matrix $a$ in the chart $\chi $, to a second one $(u'_\beta )$, having matrix $a'$ in the chart $\chi' $---so that $u'_\beta =L^\alpha  _{\ \,\beta  } \,u_\alpha $. From $u'_\beta =a'^\nu _{\ \,\beta  } \partial'_\nu $,$\quad\partial'_\nu =P^\mu _{\ \,\nu  } \partial _\mu $, and $\ \partial _\mu =b^\alpha  _{\ \,\mu }u_\alpha $ (where $b\equiv a^{-1}$), we find immediately [independently of the validity of Eqs. (\ref{purely-spatial-change}) and (\ref{P = 1 bloc 3 x 3})]:
\be\label{L for two tetrads}
L^\alpha  _{\ \,\beta  } =b^\alpha  _{\ \,\mu  }\,P^\mu _{\ \,\nu  }\,a'^\nu _{\ \,\beta  }, \qquad \mathrm{or}\quad L=b\,P\,a'.
\ee
If both tetrads are orthonormal, $L$ given by (\ref{L for two tetrads}) is automatically a Lorentz transformation. In a general spacetime, $b$ and $a'$ depend on the common time coordinate $x^0=x'^0$ as do $G$ and $G'$, while $P$ does not depend on it by virtue of (\ref{purely-spatial-change}). It is a priori unlikely that the time-dependences of $b$ and $a'$ cancel one another in Eq. (\ref{L for two tetrads}), so we expect that in general $L$ depends on $x^0$.\\

\vspace{3mm}
Let us begin to check this when, more specifically, each of the two tetrads $(u_\alpha )$ and $(u'_\beta )$ is the (orthonormal) ``Cholesky tetrad" in the chart $\chi $ (respectively $\chi '$). Thus, $a \equiv C^{-1}$ and $a' \equiv C'^{-1}$, where $C$ (respectively $C'$) is the unique lower triangular matrix with (strictly) positive diagonal entries such that $C^T \eta C = G$ (respectively $C'^T\eta C' = G'$), $G$ and $G'$ being the metric's matrices in the charts $\chi $ and $\chi '$, respectively. Note first that $a$ and $a'$ are lower triangular matrices, as are $C$ and $C'$. That is, $a^\mu _{\ \,\alpha }=0$ if $\mu < \alpha $. It follows that $a^\mu _{\ \,\mu  }=1/C^\mu _{\ \,\mu  }$ at fixed $\mu $ (without sum), so that $a$ (and $a'$) has (strictly) positive diagonal entries, as has $C$ (and $C'$). In Eq. (\ref{L for two tetrads}), we have now $C=a^{-1}$ in the place of $b$. Since $C$ and $a'$ are lower triangular matrices, (\ref{L for two tetrads}) rewrites as 
\be\label{C and a' lowerT}
L^\alpha   _{\ \,\beta  }  = \sum_{\mu =0} ^{\alpha } C^\alpha   _{\ \,\mu} \sum_{\nu =\beta } ^{3} P^\mu   _{\ \,\nu }\,a'^\nu  _{\ \,\beta  }.
\ee
By using (\ref{P = 1 bloc 3 x 3}), this  specializes, on one hand, to
\be\label{L^0_beta}
L^0  _{\ \,\beta  }  = C^0  _{\ \,0}\sum_{\nu =\beta } ^{3}  P^0   _{\ \,\nu }\,a'^\nu  _{\ \,\beta  }\quad \Rightarrow L^0  _{\ \,0 }  = C^0  _{\ \,0}\,a'^0  _{\ \,0 } \quad \mathrm{and}\ L^0  _{\ \,p }  = 0\ (p=1,2,3),
\ee
thus also $(L^{-1})^0  _{\ \,p }  = 0\ (p=1,2,3)$, from which, using $L^T=\eta L^{-1}\eta^{-1} $, we find $L^p  _{\ \,0 } = 0$. Entering this into $L^T\eta L=\eta$ gives us $(L^0  _{\ \,0 })^2  = 1$, hence, in view of $L^0  _{\ \,0 }  = C^0  _{\ \,0}\,a'^0  _{\ \,0 } >0$, $L^0  _{\ \,0 }  = 1$. Thus we have simply:
\be\label{L^0_beta-2}
L^0  _{\ \,0 }  = 1 \quad \mathrm{and}\ L^0  _{\ \,p }  = L^p  _{\ \,0 } = 0\ (p=1,2,3).
\ee
On the other hand, for the purely spatial components of $L$, (\ref{C and a' lowerT}) becomes, using again (\ref{P = 1 bloc 3 x 3}):
\be\label{L^p_q}
L^p  _{\ \,q }  = \sum_{k=1} ^{p} C^p  _{\ \,k} \sum_{j =q} ^{3} P^k   _{\ \,j}\,a'^j _{\ \,q } \qquad (p,q=1,2,3).
\ee
In particular, for $q=3$, we get
\be\label{L^p_3}
L^p  _{\ \,3 }  = a'^3 _{\ \,3 } \sum_{k=1} ^{p} C^p  _{\ \,k}  P^k   _{\ \,3} \qquad (p=1,2,3).
\ee
Moreover, since $C$ is lower triangular and verifies $C^T \eta C = G$, we have
\be
\sum_{p=j} ^{3} C^p  _{\ \,j}\,C^p  _{\ \,k}=-g_{jk} \quad (j,k=1,2,3),
\ee
so that $c^T\,c=-g$, with $c\equiv (C^p_{\ \,j})$, is the standard Cholesky decomposition of the positive definite symmetric $3\times 3$ matrix $\Mat{h}\equiv (-g_{jk})$. The same applies with primes, of course. In particular:
\be
\frac{1}{a'^3 _{\ \,3 }}=C'^3  _{\ \,3}=\sqrt{-g'_{33}}= \sqrt{-P^j   _{\ \,3} \,P^k   _{\ \,3} \,g_{jk}}\ >0.
\ee
(Of course, unless explicitly mentioned otherwise, there is summation over repeated indices, here the spatial coordinate indices $j$ and $k$ varying from $1$ to $3$.) Hence from (\ref{L^p_3}):
\be\label{L^p_3-2}
L^p  _{\ \,3 }  = \frac{\sum_{k=1} ^{p} C^p  _{\ \,k}  P^k   _{\ \,3}}{\sqrt{-P^j   _{\ \,3} \,P^k   _{\ \,3} \,g_{jk}}}  \qquad (p=1,2,3).
\ee
Clearly, this depends in general on the time coordinate as do the $C^p  _{\ \,k}$ 's and the $g_{jk}$ 's. \\

Even more specifically, let us assume that the metric has the diagonal form (\ref{diag G}) in the chart $\chi$ (but not necessarily in the chart $\chi '$). Thus
\be\label{C diag}
C=\mathrm{diag}(\sqrt{\abs{d_\mu }}),
\ee
so Eq. (\ref{L^p_3-2}) becomes
\be\label{L^p_3-3}
L^p  _{\ \,3 }  = \frac{ P^p   _{\ \,3}\,\sqrt{-d_p}}{\sqrt{-(P^j   _{\ \,3})^2 \,d_{j}}}  \qquad (\mathrm{no\ sum\ on\ }p=1,2,3).
\ee
Since $P$ does not depend on $x^0$, we find easily from this that
\be\label{L^p_3-4}
\frac{\partial }{\partial x^0}\left(L^p  _{\ \,3 }\right)  \propto P^p   _{\ \,3} (P^j   _{\ \,3})^2 \frac{\partial }{\partial x^0}\left(\frac{d_j}{d_p}\right) \quad (\mathrm{no\ sum\ on\ }p=1,2,3),
\ee
with a non-zero proportionality factor. In general, the eigenvalues $d_j$, and their ratios as well, depend on $x^0$, hence $\frac{\partial }{\partial x^0}\left(L^p  _{\ \,3 }\right)$ will be non-zero. Two natural exceptions are: i) $d_j=\omega_j({\bf x})\varphi (x^0)\quad [{\bf x}\equiv (x^j)]$---which seems however difficult to get---and ii) $d_j=d_j^0\, h(x^0,{\bf x})$ with $d_j^0$ constant ($d_j^0 < 0$ with $h > 0$). In the latter case, we get $d'_j=-h $ ($j=1,2,3$) after the coordinate change $x'^j=x^j\,\sqrt{ -d_j^0}$. That is, Case  (ii) corresponds precisely with the metric (\ref{space-isotropic diagonal}). \\

Thus, in a non-stationary spacetime, the local Lorentz transformation $L$ which transforms the Cholesky tetrad $(u_\alpha )$ got in a chart $\chi $ to the Cholesky tetrad $(u'_\alpha )$ got in a chart $\chi' $ that belongs to the same reference frame, in general depends on the time coordinate. [$L$ is given by Eqs. (\ref{L^0_beta-2}) and (\ref{L^p_q}), and this dependence is particularly obvious on Eq. (\ref{L^p_3-2}) and its specialization for the case that $G$ is diagonal, Eq. (\ref{L^p_3-4}).] The fields of Dirac matrices $\gamma ^\mu $ and $\widetilde{\gamma }^\mu $ associated respectively with the two tetrad fields $(u_\alpha )$ and $(u'_\alpha )$, Eq. (\ref{flat-deformed}), are related together by Eq. (\ref{similarity-gamma}), where $S$ is a local gauge transformation obtained by ``lifting" the local Lorentz transformation $L$. That is, $S$ is such that
\be\label{L=Lambda o S}
(\forall X \in \mathrm{V}) \quad \Lambda (S(X)) =L(X), 
\ee
where $\Lambda: {\sf Spin(1,3)} \rightarrow {\sf SO(1,3)}$ is the two-to-one covering map of the special Lorentz group by the spin group. In other words, $S$ is such that
\be\label{S=S(L)}
S(X)=\pm {\sf S}(L(X)),
\ee
where $L\mapsto \pm {\sf S}(L)$ is the spinor representation (which is defined only up to a sign, contrary to the covering map of which it is the ``inverse") \cite{A42}. (We assume that a {\it smooth} lifting is possible, i.e., that $S$ depends smoothly on $X \in \mathrm{V}$; this is always possible in a simply-connected spacetime \cite{Isham1978}.) Since, as we have shown, the local Lorentz transformation $L$ in general depends on the time coordinate, so also does the local gauge transformation (\ref{S=S(L)}). It follows that the Dirac Hamiltonian operators $\mathrm{H}$ and $\widetilde{\mathrm{H}}$ associated, in the chart $\chi $, with the fields of Dirac matrices $\gamma ^\mu $ and $\widetilde{\gamma }^\mu $, are in general not physically equivalent---at least for DFW, see Eq. (\ref{partial_0 S=0}). However, both $\mathrm{H}$ and $\widetilde{\mathrm{H}}$ are got by applying the ``Cholesky prescription" $a\equiv C^{-1}$ in two different charts {\it belonging to the same reference frame}. Therefore, that prescription is not unique.\\

\section{Case of a space-isotropic diagonal metric}\label{f -h -h -h}

As we have just seen, the ``Cholesky prescription" $a\equiv C^{-1}$ does not in general provide a unique Dirac Hamiltonian operator $\mathrm{H}$ (nor a unique energy operator $\mathrm{E}$) in a given reference frame, even if the metric is diagonal---in which case that prescription coincides with the ``diagonal tetrad" prescription (\ref{diagonal-tetrad}), which is thus non-unique in general. Therefore, we are led to consider a special form of a diagonal metric. For the reason that appeared after Eq. (\ref{L^p_3-3}), we are in fact led to consider the metric (\ref{space-isotropic diagonal}). And indeed, for this metric, the ``diagonal tetrad" prescription (\ref{diagonal-tetrad}) {\it is} unique:

\paragraph{Theorem 1.}\label{Theorem1} {\it Let the metric have the space-isotropic diagonal form (\ref{space-isotropic diagonal}) in some admissible chart $\chi : X \mapsto (x^\mu )$. Let $\chi' : X \mapsto (x'^\mu )$ be an admissible chart that defines the same reference frame $\mathrm{R}$, i.e., $\chi $ and $\chi '$ are defined on the same domain $\mathrm{U}$ and the change from $(x^\mu )$ to $ (x'^\mu )$ is purely spatial, Eq. (\ref{purely-spatial-change}).} 

(i) {\it Assume that the metric have the same} form {\it (\ref{space-isotropic diagonal}) in the new chart. Apply the ``diagonal tetrad" prescription (\ref{diagonal-tetrad}) in each of the two charts, thus getting two tetrad fields $(u_\alpha )$ and $(u'_\beta )$. Then, the tetrad fields $(u_\alpha )$ and $(u'_\beta )$ are related together by a} time-independent {\it Lorentz transformation $L$, hence give rise,} in the reference frame $\mathrm{R}$, {\it to equivalent Hamiltonian operators as well as to equivalent energy operators---for the standard version of the covariant Dirac equation, DFW.} 

(ii) {\it In addition, assume that the spatial part of the chart $\chi $ maps its domain $\mathrm{U}$ over the entire space ${\sf R}^3$. That is, $P_S: {\bf X}\equiv (x^\mu ) \mapsto {\bf x}\equiv (x^j)$ being the spatial projection ${\sf R}^4\rightarrow {\sf R}^3$, assume that $\Omega \equiv P_S(\chi(\mathrm{U}))={\sf R}^3$. Then, the Lorentz transformation $L$ that transforms the tetrad field $(u_\alpha )$ into $(u'_\beta )$ is} constant, {\it hence $(u_\alpha )$ and $(u'_\beta )$ give rise,} in any reference frame $\mathrm{F}$, {\it to equivalent Hamiltonian operators as well as to equivalent energy operators---for the DFW and for the QRD--0 versions of the Dirac equation. }\\

\noindent We begin with a lemma that will be used for the proofs of both (i) and (ii).

\paragraph{Lemma.}\label{Lemma} {\it Assume that the metric has the form (\ref{space-isotropic diagonal}) in the chart $\chi : X \mapsto (x^\mu )$. Let $\chi' : X \mapsto (x'^\mu )$ be a chart defined on the same domain $\mathrm{U}$, and assume that the change from $(x^\mu )$ to $ (x'^\mu )$ verifies Eq. (\ref{purely-spatial-change}). Define the spatial tensor $F$, with components
\be
F^j _{\ \,k  }= \frac{\partial x'^j }{\partial x^k }.
\ee
In order that the metric have the same form (\ref{space-isotropic diagonal}) in the chart $\chi '$, it is necessary and sufficient that there exist a positive function defined over the spatial range of the chart $\chi $, $\alpha : P_S(\chi (\mathrm{U}))\equiv \Omega \rightarrow {\sf R}^*_+,\quad {\bf x} \mapsto \alpha ({\bf x})$, such that}
\be\label{C=alpha^2 Id}
\mathcal{C}_R\equiv F^T\, F = \alpha ^2 \, {\bf 1}_3. 
\ee

\vspace{2mm}
\noindent {\it Proof.} Since the coordinate change from $(x^\mu )$ to $ (x'^\mu )$ is purely spatial, the matrix $P$, with components $P^\mu _{\ \,\nu  }\equiv \frac{\partial x^\mu }{\partial x'^\nu }$, has the form (\ref{P = 1 bloc 3 x 3}), thus it has the block structure
\be\label{P = 1 bloc 3 x 3-2}
P=\begin{pmatrix}
1  & 0\\ 
 0 &  Q 
\end{pmatrix},
\ee
where $Q$ is the $3\times 3$ matrix with components $Q^j _{\ \,k  }\equiv \frac{\partial x^j }{\partial x'^k }$. In the chart $\chi$, the metric has the space-isotropic diagonal form (\ref{space-isotropic diagonal}), thus in particular the metric's matrix $G$ has the same block structure as has $P$. Moreover, for $G$, the $3\times 3$ matrix which corresponds to $Q$ in Eq. (\ref{P = 1 bloc 3 x 3-2}) is just $-h{\bf 1}_3$. Hence, the metric's matrix $G'$ in the new chart still has that block structure and writes explicitly:
\be
G' = P^T\, G\,P = \begin{pmatrix}
f\  & 0\\ 
 0 &  -h\,Q^T\, Q 
\end{pmatrix}.
\ee 
It thus has also the form (\ref{space-isotropic diagonal}), iff there is some function $h'=h'((x'^\mu ))>0$ [the new coefficient for the spatial part of the metric (\ref{space-isotropic diagonal})], such that
\be
h\,Q^T\, Q = h'{\bf 1}_3,  
\ee
or equivalently, if and only if
\be\label{C=h/h' Id}
\mathcal{C}_R\equiv F^T\, F = \frac{h}{h'}{\bf 1}_3,\qquad F\equiv Q^{-1}, \quad F^j _{\ \,k  }= \frac{\partial x'^j }{\partial x^k }. 
\ee
Recall that the coordinate change is assumed purely spatial, so $F$ and $\mathcal{C}_R$ depend only on ${\bf x}\equiv (x^j) \in \Omega $. Hence, even though $h$ depends in general on the time coordinate $x^0$ as well as on the spatial coordinates $x^j$, the ratio $h/h'>0$ depends only on ${\bf x} \in \Omega $ if Eq. (\ref{C=h/h' Id}) is verified. Thus in Eq. (\ref{C=h/h' Id}) we may set $h/h'=\alpha ^2$ with $\alpha =\alpha ({\bf x}), \ {\bf x}\in \Omega $. This proves the \hyperref[Lemma]{Lemma}. \\

\noindent {\it Proof of Theorem 1.} Let us prove Point (i). As we saw at the end of Sect. \ref{Tetrad construc}, the ``diagonal tetrad" (\ref{diagonal-tetrad}) is just the Cholesky tetrad corresponding to the case of a diagonal metric. Since by assumption the metric has the space-isotropic diagonal form (\ref{space-isotropic diagonal}) in the first chart $\chi$, we thus may use Eqs. (\ref{L^0_beta-2}) and (\ref{L^p_q}), with the matrix $C$ being given by (\ref{C diag}), which becomes more particularly 
\be\label{C diag iso}
C=\mathrm{diag}\left(\sqrt{f},\ \sqrt{h},\ \sqrt{h},\ \sqrt{h}\right).
\ee
If the metric has the same form (\ref{space-isotropic diagonal}) in the new chart $\chi '$, then the corresponding diagonal tetrad $(u'_\beta )$, Eq. (\ref{diagonal-tetrad}), is such that
\be\label{a' diag iso}
(a'^j_{\ \, k})=\mathrm{diag}\left(1/\sqrt{h'},\ 1/\sqrt{h'},\ 1/\sqrt{h'}\right).
\ee
With (\ref{C diag iso}) and (\ref{a' diag iso}), Eq. (\ref{L^p_q}) becomes:
\be
L^p_{\ \,q}=P^p_{\ \,q}\ \sqrt{\frac{h}{h'}}\equiv P^p_{\ \,q}\, \alpha \qquad (p,q=1,2,3).
\ee
Here $P^p _{\ \,q} \equiv  \partial x^p/\partial x'^q$ is independent of the common time coordinate $x'^0=x^0$, as it results from (\ref{purely-spatial-change}). The same is true of $\alpha \equiv \sqrt{h/h'}$, because the assumptions of the \hyperref[Lemma]{Lemma} are satisfied. Together with (\ref{L^0_beta-2}), this means that the Lorentz transformation $L$ that takes the diagonal tetrad $(u_\alpha )$ associated with the chart $\chi $ to the diagonal tetrad $(u'_\beta )$ associated with the chart $\chi '$ is time-independent. The local gauge transformation $S$ obtained by ``lifting" that transformation $L$, Eq. (\ref{S=S(L)}), is hence time-independent also: $\partial_0  S=0$. It follows, as we noted after Eq. (\ref{SEtilde=ES modified}), that---for DFW---both the Hamiltonian and the energy operator of the reference frame R stay invariant in the sense of Eq. (\ref{similarity-invariance-H}) after the gauge transformation $S$. This proves point (i).\\

The proof of point (ii) is based on a mathematical result that pertains to continuum mechanics, namely it uses the theory of the compatibility of a field of large deformations. Consider a continuous medium whose ``initial configuration" is described by some open domain $\Omega$ of ${\sf R}^3$, and let $\mathcal{C}_R$  be a candidate field for the ``right Cauchy-Green strain tensor" of that continuous medium---thus $\mathcal{C}_R$ is a symmetric, positive definite, spatial tensor field of the $(0\ 2)$ type, defined over the domain $\Omega $. The compatibility problem is the question whether there exists a transformation $\varphi : \Omega \rightarrow {\sf R}^3, \ {\bf x}\equiv (x^j) \mapsto {\bf x}'\equiv (x'^j)$, which admit this tensor field $\mathcal{C}_R$ as its Cauchy-Green strain tensor, i.e., which be such that the gradient tensor $F$ of the transformation, with $F^j _{\ \,k  }\equiv \frac{\partial x'^j }{\partial x^k }$, verify indeed $ F^T\, F=\mathcal{C}_R$. A tractable answer to this question has been given by Vall\'ee and collaborators \cite{Vall'ee1992, Hamdouni2000, Fortun'eVall'ee2001}. Now the \hyperref[Lemma]{Lemma} above shows that the purely spatial coordinate changes that leave the form (\ref{space-isotropic diagonal}) invariant are {\it identical} with the {\it smooth} transformations $\varphi : {\bf x} \mapsto {\bf x}'=\varphi ({\bf x})=(\varphi ^j({\bf x}))$ defined over the open domain $\Omega \equiv P_S(\chi (\mathrm{U}))$ in ${\sf R}^3$ and whose associated Cauchy-Green strain tensor $\mathcal{C}_R= F^T\, F$ has the spherical form 
\be\label{C spherical}
\mathcal{C}_R({\bf x})=\alpha ({\bf x})^2\, {\bf 1}_3.
\ee
It turns out that this particular compatibility problem has been solved completely by Fortun\'e \cite{TheseFortune}: the problem can be solved iff the function $\alpha ({\bf x})>0$ {\it either } ({\bf 1}) is a positive constant $\alpha _0$ {\it or} ({\bf 2}) has the form
\be\label{alpha inversion}
\alpha ({\bf x})=\frac{b}{\abs{{\bf x-a}}^2},
\ee
where $b$ is a positive constant and ${\bf a}  \in {\sf R}^3$ is a constant vector. In case ({\bf 1}), the transformations $\varphi $ whose associated tensor $\mathcal{C}_R$ is given by (\ref{C spherical}) have the form
\be\label{phi type 1}
\varphi({\bf x})={\bf c} + \alpha _0 R.{\bf x},
\ee
where ${\bf c}  \in {\sf R}^3$ is a constant vector and $R \in {\sf O(3)}$ is a constant orthogonal transformation. 
\footnote{\label{O vs SO}\ 
In continuum mechanics, $R$ must be specifically a rotation, but this does not play any role in the derivation \cite{TheseFortune}.
}  
In case ({\bf 2}), the transformations $\varphi $ whose associated tensor $\mathcal{C}_R$ is given by (\ref{C spherical}) have the form
\be\label{phi type 2}
\varphi({\bf x})={\bf c} + b R.\frac{{\bf x-a}}{\abs{{\bf x-a}}^2},
\ee
where ${\bf c}  \in {\sf R}^3$ is a constant vector and $R \in {\sf O(3)}$ is a constant orthogonal transformation (see Note \ref{O vs SO}). Obviously, that transformation is singular as ${\bf x}\rightarrow {\bf a}$ (the image ${\bf x'}=\varphi({\bf x})$ diverges), thus it is not even defined at ${\bf x}={\bf a}$, much less smooth. However, a purely spatial coordinate change ${\bf x}'=\varphi ({\bf x})$ has to be defined and smooth on the whole of the spatial range of the first chart, thus on the whole of $\Omega \equiv P_S(\chi (\mathrm{U}))$. Therefore, the point ${\bf a}$ cannot belong to $\Omega $, so that the transformations of the type ({\bf 2}) are forbidden if $\Omega ={\sf R}^3$. It follows that only the transformations of type ({\bf 1}) are allowed. Thus the function ${\bf x}\mapsto \alpha ({\bf x})$ is a constant $\alpha _0$ and the transformation has the form (\ref{phi type 1}).\\

Point (ii) now follows easily. First note that, applying a mere homothecy-translation to the spatial part of a coordinate system in which the metric is diagonal, one leaves the diagonal tetrad (\ref{diagonal-tetrad}) unchanged: 
\be
u'_\mu \equiv \partial '_\mu /\sqrt{\abs{d'_\mu }}=\partial _\mu /\sqrt{\abs{d_\mu }}\equiv u_\mu \quad \mathrm{(no\ sum)}.
\ee
Hence, we may assume that $\alpha_0 =1$ and ${\bf c} ={\bf 0}$ in Eq. (\ref{phi type 1}). Thus $F=R$ in Eq. (\ref{C=alpha^2 Id}), i.e., $\partial x'^j/\partial x^k=R^j_{\ \, k}$, and $h'=h$ [so the metric (\ref{space-isotropic diagonal}) is left {\it invariant} by the coordinate transformation].  Hence by (\ref{diagonal-tetrad}):
\be
u_k\equiv \frac{\partial _k}{\sqrt{-d_k}}=\frac{R^j_{\ \, k}\partial '_j}{\sqrt{h}}=R^j_{\ \, k} u'_j \quad \mathrm{(no\ sum\ on\ }k=1,2,3).
\ee
Together with (\ref{L^0_beta-2}), this means that one goes from the diagonal tetrad associated with the first coordinate system to the diagonal tetrad associated with the second coordinate system by the constant Lorentz transformation
\be\label{L=1 tensor R}
L=\begin{pmatrix} 
1 & 0  & 0 & 0\\
0  &  &  & \\
0 &  & R^{-1} &\\
0 &   &  &
\end{pmatrix}.\
\ee
The local gauge transformation (\ref{S=S(L)}) obtained by ``lifting" that constant $L$ is of course constant: $\partial_\mu  S=0$. (This is true in any chart, whether it belongs to the frame $\mathrm{R}$ or not, since, as we noted after Eq. (\ref{D_mu S}), $S$ behaves as a scalar under a coordinate change.) As we concluded at the end of Section \ref{equivalent operators}, this is a sufficient condition in order that the gauge transformation $S$ lead to a Hamiltonian operator equivalent to the starting one, and also to an energy operator equivalent to the starting one, for both DFW and QRD--0. This completes the proof of \hyperref[Theorem1]{Theorem 1}.\\

\noi {\it Remarks.} {\bf i}) Assume that there exists a global chart $\chi : \mathrm{V}\rightarrow {\sf R}^4$, with moreover $\chi (\mathrm{V})={\sf R}^4$, such that the ``physical" spacetime metric $\Mat{g}$ has the form (\ref{space-isotropic diagonal}). Then of course we have $P_S(\chi (\mathrm{V}))={\sf R}^3$, so Point (ii) of \hyperref[Theorem1]{Theorem 1} applies [as well as Point (i)]. In that case, we may also define another metric on $\mathrm{V}$, say $\Mat{\gamma }$, by setting
\be
\gamma _{\mu \nu }=\eta _{\mu \nu }\equiv \mathrm{diag}(1,-1,-1,-1)
\ee
in the chart $\chi $, so that $(\mathrm{V},\Mat{\gamma })$ is the Minkowski spacetime. An obvious subcase of that particular case of validity of the whole Theorem 1 is the case $\Mat{g}=\Mat{\gamma }$, i.e., a Minkowski spacetime without gravitation. \\

\noi {\bf ii}) In the foregoing versions of the present paper, it had been found that the transformations $\varphi $ whose associated deformation tensor $\mathcal{C}_R$ is spherical, Eq. (\ref{C spherical}), were only the transformations having the form (\ref{phi type 1}), hereabove called of type ({\bf 1}). Thus, the possibility of the transformations having the form (\ref{phi type 2}), i.e., of type ({\bf 2}), had not been detected. That second possibility [which is excluded under the additional assumption made for Point (ii) of \hyperref[Theorem1]{Theorem 1}] was mentioned to the author by C. Vall\'ee (private communication). Then it was found by the author that this was due to a mistake in Eq. (61) of the former versions of this paper. Namely, although the curl of a vector field ${\bf v}$ can indeed be calculated as a vector product: $\mathrm{rot\,}{\bf v}= \nabla \wedge {\bf v}$, the curl of a vector product cannot be calculated as a double vector product.

\section{Discussion}\label{Discussion}

{\bf i}) {\it The proposed prescription and its uniqueness.} Let us assume that the metric has the space-isotropic diagonal form (\ref{space-isotropic diagonal}) in some coordinate system (chart) $\chi: X\mapsto (x^\mu )$. [One may assume that $\chi $ is a global chart, for the simplicity of discussion.] That chart defines a preferred reference frame, say R: this is the class of the charts that exchange with $\chi$ by a purely spatial coordinate change (\ref{purely-spatial-change}). We define the field of Dirac matrices in the chart $\chi$ from the diagonal tetrad field $(u_\alpha )$ defined in that same chart $\chi $. That is [Eqs. (\ref{diagonal a}) and (\ref{flat-deformed})]:
\be
\gamma ^\mu = \gamma ^{\natural \mu }/\sqrt{ \abs{d_\mu} } \quad \mathrm{(no\ sum)}, \qquad d_0\equiv f, \quad d_1=d_2=d_3 \equiv -h,
\ee
where $(\gamma ^{\natural \alpha })_{\alpha =0,...,3}$ is a set of constant ``flat" matrices, i.e., obeying the anticommutation relation (\ref{Clifford-flat-2}) with the Minkowski metric. In an arbitrary reference frame F, the field of Dirac matrices, say $\zeta ^\mu $, is defined by transforming it as a four-vector \cite{BrillWheeler1957+Corr, A45} from the chart $\chi \in \mathrm{R}$ to an arbitrary chart $\phi: X\mapsto (y^\mu ) $ belonging to the frame F: 
\be\label{gamma under chart change}
\zeta ^\mu =\frac{\partial y^\mu }{\partial x^\nu } \gamma ^\nu .
\ee
The two fields $\gamma ^\mu $ and $\zeta ^\mu $ are the expressions in the charts $\chi $ and $\phi $, respectively, of a unique ``$\gamma$ field" (Note \ref{gamma}): the one associated with the tetrad field $(u_\alpha )$ \{Ref. \cite{A45}, Appendix\}. [The fields $(u_\alpha )$ and $\gamma $ are global if the chart $\chi $ is global.] Thus, for each choice of a chart $\chi $ that both belongs to the reference frame R and is such that the metric has the form (\ref{space-isotropic diagonal}), we have one preferred tetrad field $(u_\alpha )$ and one preferred $\gamma$ field. Point (i) of \hyperref[Theorem1]{Theorem 1} proves that the choice of a such chart $\chi $ has no influence on the Hamiltonian and energy operators in the reference frame $\mathrm{R}$. Point (ii) proves that the choice of a such chart $\chi $, if its spatial range is $\Omega ={\sf R}^3$, has no influence on these operators in any reference frame $\mathrm{F}$. By the way, note that the invariance of the similarity matrix $S$ under a coordinate change, used at the end of the proof of that theorem, is easily rechecked from (\ref{similarity-gamma}) and (\ref{gamma under chart change}). \\

Of course, besides changing the chart in the particular reference frame R, one may also change the chart in the general reference frame F, say from $\phi: X\mapsto (y^\mu )$ to $\phi': X\mapsto (y'^\mu )$. But this does not change the $\gamma $ field. In other words, the two fields of Dirac matrices $\zeta ^\mu $ and $\zeta'^\mu $ got by Eq. (\ref{gamma under chart change}) are related together by the 4-vector transformation: $\zeta'^\mu =\frac{\partial y'^\mu }{\partial y^\nu } \zeta ^\nu$, with a transformation $(y^\mu ) \mapsto (y'^\mu )$ that is purely spatial---so that, as recalled in Section \ref{equivalent operators}, the Hamiltonian operator, as well as the energy operator, remain the same \cite{A42,A47}. \\ 

The choice of the set of constant ``flat" matrices, i.e. the set $(\gamma ^{\natural \alpha })_{\alpha =0,...,3}$ in Eq. (\ref{flat-deformed}), has no influence either. Indeed, two different choices, say $(\gamma ^{\natural \alpha })$ and $(\gamma ^{\sharp \alpha })$, give by (\ref{flat-deformed}) two fields of Dirac matrices which exchange by the same constant similarity that transforms $(\gamma ^{\natural \alpha })$ to $(\gamma ^{\sharp \alpha })$.\\ 

All of these results are valid for both the standard version of the curved-spacetime Dirac equation (``DFW") and the alternative version with zero connection matrices (``QRD-0"). These two versions are in general not equivalent: although the DFW equation is equivalent to a particular case of the QRD-0 equation \cite{A46}, this equivalence takes place after changing the field of Dirac matrices in a way that has no reason to be compatible with our prescription. Thus, our prescription leaves us with two non-equivalent choices for the Dirac equation in a given spacetime $(\mathrm{V},\Mat{g})$, leading to two non-equivalent choices for the Dirac Hamiltonian and for the Dirac energy operator in a given reference frame.
\footnote{\ 
Two points may further be noted in this connection. {\bf i}) Not only a unique DFW equation (as is usually the case) but also a unique QRD--0 equation is got by the proposed prescription. Indeed, two different choices of the chart $\chi \in \mathrm{R}$ (each with spatial range $\Omega ={\sf R}^3$) lead to $\gamma $ fields exchanging by a constant similarity transformation $S$: $\partial_\mu S=0$. Now, since the QRD--0 equation fixes the connection matrices $\Gamma _\mu $ (to zero), the condition of covariance of the Dirac equation under a similarity is derived in the same way as in Ref. \cite{A42}, and is thus the same \{Eq. (69) in Ref. \cite{A42}\}, i.e.: $S^\dagger B ^\mu (D_\mu S) = (D_\mu S^\dagger) B^\mu S$, where $B^\mu \equiv A\gamma ^\mu $. Thus, with $\Gamma _\mu =0$, this is verified by $S$ since $\partial_\mu S=0$. {\bf ii}) Any form of the 4-scalar Dirac equation (``QRD") is equivalent to a 4-vector Dirac equation (``TRD"), essentially by switching from the global frame field $(E_a)$ on $\mathrm{V} \times {\sf C}^4$ to an (arbitrary) global frame field $(e_a)$ on $\mathrm{T}_{\sf C}\mathrm{V}$ (\cite{A45}, Theorem 1). This TRD equation, and the Hamiltonian and energy operators associated with it in a given reference frame, have identical local expressions with those of, respectively, the starting QRD equation and the corresponding $\mathrm{H}$ and $\mathrm{E}$ operators. Therefore, the two non-equivalent choices for the Dirac equation in a given spacetime, got by using our prescription, can be regarded as pertaining to TRD as well as to QRD. In short, we have two choices only, regardless of the four-scalar or four-vector representation. 
}
Experiment might decide between the two choices.\\

\noi {\bf ii}) {\it Physical relevance of the metric (\ref{space-isotropic diagonal}).} The spacetime metric cannot generically be put in the form (\ref{space-isotropic diagonal}) in a coordinate chart for general relativity. Indeed, even the standard post-Newtonian (PN) metric, which is valid for the usual, weak gravitational fields, has non-diagonal terms: specifically it has terms $g_{0j} \ (j=1,2,3)$. However, apart from these terms it has just the form (\ref{space-isotropic diagonal}), moreover the $g_{0j}$ terms are $O(c^{-3})$ and are very small as compared with the leading corrections to the Minkowski metric, which are $O(c^{-2})$ and depend on the Newtonian potential \cite{Fock64,Weinberg}. The energy levels of a Dirac particle, as obtained with the $O(c^{-2})$ corrections to the Minkowski metric, differ already very little from the energy levels got with the non-relativistic Schr\"odinger equation in the Newtonian potential \cite{A38,Boulanger-Spindel2006}. In addition, the metric (\ref{space-isotropic diagonal}), assumed valid in the reference frame R, transforms to a more general form with non-zero $g_{0j}$ terms, after a boost and/or a rotational motion. If these involve linear velocities of post-Newtonian order: $v=O(c^{-1})$, then one gets $g_{0j}=O(c^{-3})$ as in the standard PN metric. Thus, the form (\ref{space-isotropic diagonal}) is general enough for the prospective purpose of testing the generally-covariant Dirac equations in a realistic spacetime metric.
\footnote{\
See Ref. \cite{A48v2} for a proof that our assumption: (A) ``There is a local chart $\chi : \mathrm{V}\supset \mathrm{U}\rightarrow {\sf R}^4$, such that the local expression of the metric $\Mat{g}$ is (\ref{space-isotropic diagonal}) in that chart," is invariant under (isometric) diffeomorphisms.  
}\\

\noi {\bf iii}) {\it More general metrics and other prescriptions?} The metric (\ref{space-isotropic diagonal}) is valid in preferred coordinate systems (indeed global ones whose spatial range is the whole of ${\sf R}^3$) in an alternative, scalar theory of gravity, and this in the most general case \cite{A35}. Within general relativity (GR), on the other hand, one may consider it frustrating that a unique prescription has been found only for that metric (\ref{space-isotropic diagonal})---even though, as shown above, this metric is general enough for the foreseeable experimental tests. However, recall that we were naturally led to this metric in Section \ref{Pb Cholesky} in order to get unique Hamiltonian and energy operators from choosing the ``diagonal tetrad". Indeed, we do not see any other general solution to get $\frac{\partial }{\partial x^0}\left(L^p  _{\ \,3 }\right)=0$ in Eq. (\ref{L^p_3-4}). Moreover, to have this is just a {\it necessary} condition in order that the Cholesky prescription be unique for some diagonal metric. For the well-known exact solutions of GR, either the metric can be put into the form (\ref{space-isotropic diagonal}), as is the case for the Schwarzschild spacetime (choosing isotropic coordinates) and the Friedmann-Lema\^itre-Robertson-Walker spacetimes, or it is unlikely that the Cholesky prescription is unique, as for e.g. the Kerr spacetime. 
One may also examine other prescriptions. There is indeed one known other prescription which provides a unique tetrad field in a given coordinate chart. This is the ``spatially symmetric time gauge" \cite{Kibble1963, SteinhoffSchafer2009}, which imposes to the tetrad's matrix $a$ two conditions: (i) $a^0_{\ \,p}=0\ (p=1,2,3)$ (``time gauge" \cite{Schwinger1963}, leading to what is called a Schwinger tetrad in Refs. \cite{GorbatenkoNeznamov2011,GorbatenkoNeznamov2011b}; see also Ref. \cite{DeWitt2003p361}); {\it and} (ii) the matrix $(a^j_{\ \,p})\ (j,p=1,2,3)$ is got from the (symmetric) positive square root, say  $\hat{a}\equiv \sqrt{\Mat{h}}$, of the positive definite symmetric matrix $\Mat{h}\equiv (-g_{jk})\ (j,k=1,2,3)$. [Note that, for a diagonal metric, thus in particular for the metric (\ref{space-isotropic diagonal}), both the Cholesky tetrad and the ``spatially symmetric time gauge" tetrad coincide with the diagonal tetrad (\ref{diagonal-tetrad}).] However, the tetrads got by applying that prescription in two admissible charts $\chi $ and $\chi '$ that belong to the same reference frame F are still related by Eq. (\ref{L for two tetrads}). From (\ref{P = 1 bloc 3 x 3}) we get $\Mat{h}'=Q^T \,\Mat{h}\,Q$, with $Q\equiv (P^j_{\ \,k})$, thus $\Mat{h}'=R^T \,U\Mat{h}U\,R$, where $Q=UR$ is the right polar decomposition. Since the eigenvectors of $U$ and $\Mat{h}$ do not coincide for a general matrix $Q$, there is no simple relation between $\sqrt{\Mat{h}}$ and $\sqrt{\Mat{h}'}$, and thus there is no simple relation between $a$ and $a'$. It is hence very unlikely that the time-dependences of $b\equiv a^{-1}$ and $a'$ may cancel one another in Eq. (\ref{L for two tetrads}) for a general-enough metric differing from (\ref{space-isotropic diagonal}). So that $L$ should depend on $x^0$, which would mean that Kibble's prescription \cite{Kibble1963, SteinhoffSchafer2009} is not unique and can hardly be hoped to be made unique except for the metric (\ref{space-isotropic diagonal}), just like the Cholesky prescription.  \\

{\bf Acknowledgement.} I am grateful to Frank Reifler for allowing me to use a private communication and for remarks about the present paper, to J. Brian Pitts for pointing out  Ref. \cite{OgievetskiiPolubarinov1965}, and to both for exchanges about the non-uniqueness problem. Thanks also to Claude Vall\'ee who pointed out the additional class of transformations (\ref{phi type 2}).

\appendix
\section{Appendix: First vs. second quantization}\label{FirstQuantized}

The non-uniqueness problem at hand has been proved to be there for the Hamiltonian and energy operators associated with the covariant Dirac equation, thus for {\it first-quantized} Dirac theory \cite{A43}. So the present proposal for a solution remains within first--quantized theory. There is a well-known argument against first-quantized Dirac theory: due to the negative-energy states, the positive-energy levels---e.g., of an atomic electron---would be instable since the system would make transitions to negative energy levels. However, it is also well known that, in low-energy situations, to which the use of the first-quantized Dirac theory is limited, the transition between positive and negative energy states can safely be ignored and consideration can safely be restricted to positive-energy states. For instance, the classical discussion of the hydrogen-type atoms uses the Dirac positive-energy states of the electron in the nucleus's electromagnetic field. At low energies, one thus may add a selection rule that only the positive-energy states are allowed. The quantum mechanics of the Dirac equation makes a scientific literature of tens of thousands of papers and books, including some of the pillars of modern physics. The energy levels of specifically the DFW equation were discussed by well-known physicists such as Brill \& Wheeler \cite{BrillWheeler1957+Corr}, Greiner et al. \cite{Greiner-et-al1983}, Hehl \& Ni \cite{HehlNi1990}, and Ryder \cite{Ryder2008}, to name just a few. Indeed, the three effects on quantum particles in the gravitational field which have been observed so far---the COW effect \cite{COW1975}, the Sagnac effect \cite{WernerStaudenmannColella1979}, and the quantization of the energy levels \cite{Nesvizhevsky2002}---are low-energy effects and are discussed in first-quantized Dirac theory. This is also true for the Mashhoon effect \cite{Mashhoon1988,HehlNi1990}, although testing the latter is still an experimental challenge. Obviously, the non-uniqueness of the first-quantized Dirac Hamiltonian operator and the energy spectrum is a serious problem for the discussion of these effects \cite{A43}. Moreover, note that the non-uniqueness problem is there in any admissible coordinate system in essentially any spacetime \cite{A43}. Thus, it is there already in a Cartesian coordinate system in a Minkowski spacetime \cite{A47}---as soon as one uses the DFW equation with its gauge freedom, instead of using the original Dirac equation. It is easy to check that this holds true in the presence of an electromagnetic field \cite{A48v2}. Therefore, the classical discussion of the hydrogen-type atoms, which is based on the quantum-mechanical Hamiltonian and energy spectrum, cannot be done if one uses the DFW equation with its gauge freedom.\\

One may ask whether, instead of restricting the admissible gauge transformations as was done here, this non-uniqueness could be solved by going to a second-quantized theory. However, a central object in quantum field theory (QFT) is the quantum energy-momentum tensor. This should be derived from the ``classical" energy-momentum tensor, i.e. in fact, that of the first-quantized Dirac theory. Even in that ``classical" theory, the definition of the energy-momentum tensor is not fully clear. The canonical definition of that tensor:
\begin{equation}\label{t^mu _nu}
t^\mu _{\ \nu } = \frac{\partial L}{\partial (\partial_\mu \Psi) } \ \partial_\nu  \Psi 
\ +\  \partial_\nu  \Psi^\dagger \ \frac{\partial L}{\partial (\partial_\mu  \Psi^\dagger)} 
\ -\  \delta^\mu _\nu  L
\end{equation}
is univoque at least when the Lagrangian $L$ is fixed, and is often used in the literature (e.g. \cite{Leclerc2006,L&L}). The following Lagrangian  \cite{A43} extends the standard DFW Lagrangian \cite{BrillWheeler1957+Corr,Leclerc2006} to the case of a general hermitizing matrix field $A$:
\be\label{Lagrangian}
L=\frac{i}{2}\left[\,\Psi ^\dagger A\gamma^{\mu}(D_{\mu}\Psi)-
\left(D_{\mu}\Psi \right)^\dagger A\gamma^{\mu}\Psi+2im\Psi ^\dagger A\Psi\right],
\ee
and the associated Euler-Lagrange equation is indeed the general Dirac equation (\ref{Dirac-general}). From (\ref{t^mu _nu}) and (\ref{Lagrangian}), we get
\begin{equation}\label{t^mu _nu-Dirac}
t^\mu _{\ \nu } = \frac{i}{2}\left[\,\Psi ^\dagger A\gamma^{\mu}(\partial _{\nu}\Psi)-
\left(\partial _{\nu}\Psi \right)^\dagger A\gamma^{\mu}\Psi \right]\ -\  \delta^\mu _\nu  L.
\end{equation}
Since $\Psi$ and $A$ transform as scalars and $\gamma ^\mu $ as a vector under a change of the chart, (\ref{t^mu _nu-Dirac}) defines a generally-covariant $(1\ 1)$ tensor. One may replace $t^{\mu \nu }$ by its symmetric part, but this is indifferent in what follows. The problem is that the ``field energy" $E$ of the ``classical" Dirac field obeying Eq. (\ref{Dirac-general}), is equal to the expected value of the energy operator $\mathrm{E}$ \cite{A43}:
\be\label{Field energy = mean value of E}
E\equiv \int t^0_{\ \,0} \sqrt{-g} \ \dd ^3{\bf x} = (\Psi \mid \mathrm{E}\Psi  )\equiv \int \,\Psi^\dagger A\gamma ^0 (\mathrm{E}\Psi)\,\sqrt{-g} \ \dd ^3{\bf x},
\ee
as was shown by Leclerc \cite{Leclerc2006} in a less general setting. [Using the fact that $L$ vanishes ``on shell", this is easy to check from (\ref{t^mu _nu-Dirac}) and the definition (\ref{E:=H^s}).] Thus, the non-uniqueness of the energy operator $\mathrm{E}$ translates to that of the ``classical" energy-momentum tensor, indeed even to the space integral of the latter. \\

On the other hand, there is also Hilbert's definition of the energy-momentum tensor, as the derivative of the Lagrangian density $\mathcal{L}=\sqrt{-g}\,L$ with respect to variations of the metric. This leads (up to the symmetrization) to just the same Eq. (\ref{t^mu _nu-Dirac}), though with the partial derivatives replaced by covariant derivatives \cite{BrillWheeler1957+Corr,Weldon2001}. It is easy to check that, in contrast to the canonical tensor [Eq. (\ref{t^mu _nu-Dirac}) as it is], that other tensor is gauge-invariant, thus unique. However, Eq. (\ref{Field energy = mean value of E}) then does not hold true (except for the non-generic case that the two tensors concide). I.e., the gauge-invariant field energy thus obtained, say $E'\ne E$, is not equal to the mean value $(\Psi \mid \mathrm{E}\Psi  )$ of the energy operator (\ref{E:=H^s}), and therefore is not relevant. (If one would define a different energy operator $\mathrm{E}'\ne \mathrm{E}$ from its mean values $E'$, this would have no relation to the Hamiltonian $\mathrm{H}$, moreover it would be unfeasible to calculate its energy spectrum.) Since the same gauge freedom should apply to the first and second quantized Dirac theories, and since the predictions of the first-quantized theory should be deducible from the second-quantized ones in the low energy limit, this means that the relevant quantum energy-momentum tensor should be got from ``quantizing" the {\it canonical} tensor---and thus should share this non-uniqueness with its classical counterpart. \\

Moreover, in the works on QFT of the Dirac field in a curved spacetime (e.g. \cite{Dimock1982,Dappiaggi2009,Sanders2010}), we do not see any outline of the computation of one of the already-measured or prospective effects mentioned above \cite{Mashhoon1988,HehlNi1990,COW1975,WernerStaudenmannColella1979,Nesvizhevsky2002}, which would allow one to hope that this effect might be unambiguously predictable in that framework. Finally, as noted above, the non-uniqueness problem found for the covariant first-quantized Dirac theory affects also the case with an electromagnetic field in a flat spacetime, even in an inertial frame. We therefore feel that one should first build an unambiguous first-quantized theory, as we attempted to do here.


\end{document}